\documentclass[preprint,11pt,3p,authoryear]{elsarticle}

\usepackage{amssymb}
\usepackage{tikz}
\usepackage{caption}
\usepackage{mathtools}
\usepackage{amsmath}
\usepackage{multirow}
\usepackage{booktabs}
\usepackage{colortbl}
\usepackage{color}
\usepackage{subcaption}
\usepackage{tabulary}
\usepackage{graphicx}
\usepackage{hyperref}
\usepackage{float,enumitem}
\usepackage{amssymb}
\usepackage{bm}
\usepackage{pifont}
\newcommand{\norm}[1]{\left\lVert#1\right\rVert}
\hypersetup{
colorlinks=true,
linkcolor=blue,
citecolor=blue,
urlcolor=blue}

\usepackage{geometry}
\usepackage{setspace}
\usepackage{kpfonts}

\journal{arXiv}

\begin{document}

\begin{frontmatter}
  \title{D\'ej\`a vu: A data-centric forecasting approach \\ through time series cross-similarity}

\author[label4]{Yanfei Kang}
\address[label4]{School of Economics and Management, Beihang University, Beijing, China}

\author[label1]{Evangelos Spiliotis}
\address[label1]{Forecasting and Strategy Unit, School of Electrical and Computer Engineering, National Technical University of Athens, Greece}

\author[label5]{Fotios Petropoulos}
\address[label5]{School of Management, University of Bath, UK}

\author[label1]{Nikolaos Athiniotis}

\author[label6]{Feng Li\corref{cor1}}
\ead{feng.li@cufe.edu.cn}
\cortext[cor1]{Correspondance: Feng Li, School of Statistics and Mathematics, Central University of Finance and Economics, Shahe Higher Education Park, Changping District, Beijing 102206, China.}
\address[label6]{School of Statistics and Mathematics, Central University of Finance and Economics, Beijing, China}

\author[label1]{Vassilios Assimakopoulos}

\begin{abstract}
Accurate forecasts are vital for supporting the decisions of modern companies. Forecasters typically select the most appropriate statistical model for each time series. However, statistical models usually presume some data generation process while making strong assumptions about the errors. In this paper, we present a novel data-centric approach --- `forecasting with similarity', which tackles model uncertainty in a model-free manner. Existing similarity-based methods focus on identifying similar patterns within the series, i.e., `self-similarity'. In contrast, we propose searching for similar patterns from a reference set, i.e., `cross-similarity'. Instead of extrapolating, the future paths of the similar series are aggregated to obtain the forecasts of the target series. Building on the cross-learning concept, our approach allows the application of similarity-based forecasting on series with limited lengths. We evaluate the approach using a rich collection of real data and show that it yields competitive accuracy in both points forecasts and prediction intervals.

\end{abstract}


\end{frontmatter}
 \newpage
\section{Introduction}
\label{sec:sec1}

Effective forecasting is crucial for various functions of modern companies. Forecasts are
used to make decisions concerning business operations, finance, strategy, planning,
and scheduling, among others. Despite its importance, forecasting is not a straightforward
task. The inherent uncertainty renders the provision of perfect forecasts
impossible. Nevertheless, reducing the forecast error as much as possible is expected to
bring significant monetary savings.

We identify the search for an ``optimal'' model as the main challenge to
forecasting. Existing statistical forecasting models implicitly assume an underlying data
generating process (DGP) coupled with distributional assumptions of the forecast errors
that do not essentially hold in practice. \cite{Petropoulos2018-cl} suggest that three
sources of uncertainty exist in forecasting: model, parameter, and data. They found that
merely tackling the model uncertainty is sufficient to bring most of the performance
benefits. This result reconfirms George Box's famous quote, ``all models are wrong, but
some are useful.'' It is not surprising that researchers increasingly avoid using a single
model, and opt for combinations of forecasts from multiple models
\citep{Jose2008,Kolassa2011,blanc2016when,Bergmeir2016-su,Petropoulos2018-mt,Monteros2019-oy}. We argue that there is
another way to avoid selecting a single model: to select no models at all.

This study provides a new way to forecasting that does not require the estimation of any
forecasting models, while also exploiting the benefits of cross-learning
\citep{Makridakis2019-oy}. With our proposed approach, a target series is compared against
a set of reference series attempting to identify similar ones (d\'ej\`a vu). The point
forecasts for the target series are the average of the future paths of the most similar
reference series.  The prediction intervals are based on the distribution of the reference
series, calibrated for low sampling variability.  Note that no model extrapolations take
place in our approach. The proposed approach has several advantages compared to existing
methods, namely (\textit{i}) it tackles both model and parameter uncertainties, (\textit{ii}) it does not
use time series features or other statistics as a proxy for determining similarity, and
(\textit{iii}) no explicit assumptions are made about the DGP as well as the distribution of the
forecast errors.

We evaluate the proposed forecasting approach using the M1 and M3 competition data
\citep{Makridakis1982111, Makridakis2000a}. Our approach results in good point forecast accuracy, which is on par with state-of-the-art statistical benchmarks, while a simple combination of our data-centric approach and exponential smoothing significantly outperforms all other approaches tested. Also,
forecasting with cross-similarity offers a better estimation of forecast uncertainty, which would
allow achieving higher customer service levels.

The rest of the paper is organized as follows. In the next section, we present an overview
of the existing literature and provide our motivation behind ``forecasting with
cross-similarity'', . Section \ref{sec:methodology} describes the methodology for the
proposed forecasting approach, while section \ref{sec:evaluation} presents the
experimental design and the results. Section \ref{sec:discussions} offers our discussions
and insights, as well as implications for research and practice. Finally, section
\ref{sec:conclusions} provides our concluding remarks.

\section{Background research} \label{sec:literature}

\subsection{Forecast model selection} \label{sec:selection_combination}

When forecasting with numerous time series, forecasters typically try to enhance
forecasting accuracy by selecting the most appropriate model from a set of
alternatives. The solution might involve either aggregate selection, where a single model
is used to extrapolate all the series, or individual selection, where the most appropriate
model is used per series \citep{Fildes1989aa}. The latter approach can provide
substantial improvements if forecasters are indeed in a position to select the best model
\citep{2001537Fildes,fildes2015simple}. Unfortunately, this is far from the reality due to the presence of
data, model, and parameter uncertainties \citep{Kourentzes2014291, Petropoulos2018-cl}.

In this respect, individual selection becomes a complicated problem and forecasters have
to balance the potential gains in forecasting accuracy and the additional complexity
introduced. Automatic forecasting algorithms test multiple forecasting models and select
the `best' based on some criterion. The criteria include information criteria,
e.g., the likelihood of a model penalised by its complexity \citep{Hyndman2002,Hyndman2008b},
or rules based on forecasting performance on past windows of the data
\citep{Tashman00}. Other approaches to model selection involve discriminant analysis
\citep{SHAH1997489}, time-series features \citep{Petropoulos2014152}, and expert rules
\citep{ADYA2001143}. An interesting alternative is to apply cross-learning so that
the series are clustered based on an array of features and the best
model is selected for their extrapolation \citep{Kang2017345, Spiliotis2019M3}.

In any case, the difference between two models might be small, and the selection of one
over the other might be purely due to chance. The small differences between models
also result in different models being selected when different criteria or cost functions are used
\citep{Billah2006-jg}. Moreover, the features and the rules considered may not be adequate
for describing every possible pattern of data. As a result, in most cases, a clear-cut for
the `best' model does not exist because all models simply are rough approximations of
the reality.

\subsection{The non-existence of a DGP and forecast model combination}\label{sec:no_dgp}

Time series models that are usually offered by the off-the-shelf forecasting software have
over-simplified assumptions (such as the normality of the residuals and stationarity),
which do not necessarily hold in practice. As a result, it is impossible for these models
to capture the actual DGP of the data perfectly. One could work towards defining a complex
multivariate model \citep{Svetunkov_undated-vo}, but this would lead to all kinds of new
problems, such as data limitations and the inability to accurately forecast some of the
exogenous variables, which are identified as significant.

As a solution to the above problem, forecasting researchers have been combining
forecasts from different models
\citep{Bates1969,Clemen1989,Makridakis1983c,Timmermann2006,Claeskens2016,blanc2016when}. The main
advantage of combining forecasts is that it reduces the uncertainty related to model and
parameter determination, and decreases the risk of selecting a single and inadequate
model. Moreover, combining different models enables capturing multiple patterns. Thus,
forecast combinations lead to more accurate and robust forecasts with lower error
variances \citep{Hibon2005}.

Through the years, the forecast combination puzzle \citep{Watson2004Combination,Claeskens2016,blanc2016when}, i.e., the fact
that optimal weights often perform poorly in applications, has been both theoretically and
empirically examined. Many alternatives have been proposed to exploit the benefits of
combination, including Akakie's weights \citep{Kolassa2011}, temporal aggregation levels
\citep{Kourentzes2014291,kourentzes2017demand}, bagging \citep{Bergmeir2016-su, Petropoulos2018-cl}, and
hierarchies \citep{Hyndman2011,ATHANASOPOULOS201760}, among others. Moreover, simple
combinations have been shown to perform well in practice \citep{Petropoulos2019SCUM}.
In spite of the improved performance offered by forecast combination, some primary
difficulties, e.g., (\textit{i}) determining the pool of models being averaged, (\textit{ii}) identifying
their weights, and (\textit{iii}) estimating multiple models, prevent forecast combination from being
widely applied by practitioners.

\subsection{Forecasting with cross-similarity}\label{sec:review_similar}

An alternative to fitting statistical models to the historical data would be exploring
whether similar patterns have appeared in the past. The motivation behind this argument
originates from the work on structured analogies by \cite{Green2007-ts}. Structured
analogies is a framework for eliciting human judgment in forecasting. Given a forecasting
challenge, a panel of experts is assembled and asked to independently and anonymously
provide a list of analogies that are similar to the target problem together with the
degree of similarity and their outcomes. A facilitator calculates the forecasts for the
target situation by averaging the outcomes of the analogous cases weighted by the degree
of their likeness.

Given the core framework of structured analogies described above, several modifications
have been proposed in the literature. Such an approach is practical in cases that no
historical data are available \citep[e.g.,][]{Nikolopoulos2015-ia}, which renders the
application of statistical algorithms impossible. Forecasting by analogy has also been
used in tasks related to new product forecast
\citep{Goodwin2013-hu,Wright2015-ji,Hu2019-sa}, in which the demand and the life-cycle
curve parameters are possible to estimate based on the historical demand values and
life-cycles of similar products.

Even when historical information is available, sharing information across series has been
shown to improve the forecasting performance. A series of studies attempted to estimate
the seasonality on a group level instead of a series level
\citep[e.g.,][]{Mohammadipour2012,Zhang2013-dg,Boylan2014-sl}. When series are arranged in
hierarchies, it is possible to have similarities in seasonal patterns among products that
belong to the same category. This renders their estimation on an aggregate level more
accurate, especially for the shorter series where few seasonal cycles are available.

The use of cross-sectional information for time series forecasting tasks is a feature of
the two best-performing approaches by \citet{Smyl2019-oy} and \citet{Monteros2019-oy} in
the recent M4 forecasting competition \citep{Makridakis2019-oy}. \cite{Smyl2019-oy}
propose a hybrid approach that combines exponential smoothing with neural networks. The
hierarchical estimation of the parameters utilises learning across series but also focuses
on the idiosyncrasies of each series. \cite{Monteros2019-oy} use cross-learning based on
the similarity of the features in collections of series to estimate the combination
weights assigned to a pool of forecasting methods.

A stream of research has focused on similarity-based approaches to forecasting. Similarity-based is
based on an assumption nicely articulated by \cite{Dudek2010}: ``If the process pattern $x_\alpha$
in a period preceding the forecast moment is similar to the pattern $x_b$ from the history of this
process, then the forecast pattern $y_\alpha$ is similar to the forecast pattern $y_b$.''
In other words, \cite{Dudek2010} suggested that similar patterns may exist within the same signal
(process), i.e., the same time series. He applied similarity-based approaches for short-term load
forecasting, and empirically demonstrated the usefulness of this approach \citep{Dudek2010,Dudek2015-yb}.
Also, \cite{Dudek2015-oo} discussed that similar patterns may be identified via a variety of methods (such as the
kernel and nearest neighbour methods). Finally, he discussed that the advantages of forecasting by similarity
include simplicity, ease of estimation and calculation, and ability to deal with missing data.

Along the same lines, \cite{Nikolopoulos2016-gr} explored the value of identifying similar patterns within a
series of intermittent nature (where the demand for some periods is zero). They proposed
an approach that uses nearest neighbours to predict incomplete series of consecutive
periods with non-zero demand values based on past occurrences of non-zero demands. \cite{Martinez2019-bw} and \cite{Martinez2019-ve} also used $k$-nearest neighbours to find similar patterns in fast-moving series and use them for extrapolation. They also suggested the use of multiple $k$ values through ensembles to tackle the need of selecting a single $k$ parameter in the nearest neighbours method.

\cite{Li2019-kk} focused on fast-moving data in the context of maritime, and suggested that
the time series is decomposed in low and high frequency components. Subsequently, they suggested
similarity grouping of overlapping segments of the high frequency component towards
producing its prediction with neural networks. \cite{Li2019-kk} used dynamic time warping (DTW) to measure similarity. DTW is an algorithm for identifying alternative alignments between the points of
two series, so that their total distance is minimized. Indeed, \cite{Li2020-kb} showed that DTW is superior to Euclidean distance in classifying and clustering time series, and it could be further improved by considering adaptive constrain. In any case, similar to the previous studies by Dudek, Nikolopoulos and Mart\'inez, the
approach by \cite{Li2019-kk} focused on self-similarities: similarities in the patterns
within a time series.

We are proposing a novel approach to forecasting that builds on existing approaches on
forecasting with cross-similarity, but also extends them in the sense that we
suggest that searching for similar patterns can be expanded from within-series to
across-series. While cross-learning information has been used in forecasting previously,
to the best of our knowledge, it has not been utilised for directly judging the similarity
of different series, without the need to extract and estimate time series features.
Directly looking for similar observed patterns in the historical information of other series
might be particularly relevant in sets of data where appropriate clusters (subsets) are characterised by
homogeneity, which could be the case in the sales or demand patterns of a
distinct category of products observed by a retailer. Cross-series similarity is also
appealing for the cases of short series, where the limited historical information does not
allow for learning through self-similarities. In any case, in searching for similarity,
it might be useful to consider a decomposition of low and high frequency components, as
suggested by \cite{Li2019-kk}.

\section{Methodology}\label{sec:methodology}

Given a set with rich and diverse reference series, the objective of forecasting with
cross-similarity is to find the most similar ones to a target series, average their future
paths, and use this average as the forecasts for the target series. We assume that the
target series, $y$, has a length of $n$ observations and a forecasting horizon of
$h$. Series in the reference set shorter than $n+h$ are not considered. Series longer than
$n+h$ are truncated, keeping the last $n+h$ values. The first $n$ values are used for
measuring similarity and the last $h$ values serve as the future paths. We end up with a
matrix $Q$ of size $m \times (n+h)$. Each row of $Q$ represents the
$n+h$ values of a (truncated) reference series, and $m$ is the number of the reference
series.  A particular reference series is denoted with $Q(i)$, where
$i \in {1, \dots, m}$, $Q(i)_{1, \dots, n}$ is the historical data, and
$Q(i)_{n+1, \dots, n+h}$ represents the future paths. The proposed approach consists of the
following steps.

\begin{enumerate}[noitemsep, leftmargin=*,labelindent=16pt, label=\bfseries Step \arabic*]
\item \textbf{Removing seasonality}, if a series is identified as seasonal.
\item \textbf{Smoothing} by estimating the trend component through time series decomposition.
\item \textbf{Scaling} to render the target and possible similar series comparable.
\item \textbf{Measuring similarity} by using a set of distance measures.
\item \textbf{Forecasting} by aggregating the paths of the most similar series.
\item \textbf{Inverse scaling} to bring the forecasts for the target series back to its original scale.
\item \textbf{Recovering seasonality}, if the target series is found seasonal in Step 1.
\end{enumerate}

In the following subsections, we describe these steps in details. Section
\ref{sec:preprocessing} describes the preprocessing of the data (Steps 1, 2, 3, 6, and 7),
section \ref{sec:similarity} provides the details regarding similarity measurement and
forecasting (Steps 4 and 5), while section \ref{sec:PI} explains how prediction intervals
are derived.

\subsection{Preprocessing}\label{sec:preprocessing}

When dealing with diverse data, preprocessing becomes essential for effectively
forecasting with cross-similarity. This is because the process of identifying similar series is
complicated when multiple seasonal patterns and randomness are present, and the scales
of the series to be compared differ. If the reference series are not representative of the
target series or the reference set is lack of diversity, the chances of observing similar
patterns are further decreased.

To deal with this problem, we consider three steps which are applied
sequentially. The first step removes the seasonality if the series is identified as
seasonal. By doing so, the target series is more likely to effectively match with multiple
reference series, at least when dissimilarities are present due to different seasonal
patterns. In the second step, we smooth the seasonally adjusted series to remove
randomness and possible outliers from the data, which further reduces the risk of
identifying too few similar series. Finally, we scale the target and the reference series to
the same magnitude, so that their values are directly comparable. The preprocessing step
is applied to both the reference and target series.

\subsubsection{Seasonal adjustment}\label{sec:seasonality}

Seasonal adjustment is performed by utilizing the ``Seasonal and Trend decomposition using
Loess'' (STL) method presented by \cite{Cleveland1990} and implemented in the
\textit{stats} package for R. In brief, STL decomposes a time series $x_t$ into the trend ($T$),
seasonal ($S$), and remainder ($R$) components, assuming additive interactions among them: $x_t = T_t + S_t + R_t$. An
adjustment is only considered if the series is identified as seasonal, through a
seasonality test. The test \citep{Assimakopoulos2000521,Fioruci2016} checks for autocorrelation significance on the $s^{\text{th}}$
term of the autocorrelation function (ACF), where $s$ is the frequency of the series (e.g., $s=12$ for monthly
data). Thus, given a series of $\hat{n}\geq3s$ observations, frequency $s>1$, and a
confidence level of 90\%, a seasonal adjustment is considered only if
\begin{equation*}
    |\text{ACF}_s| > 1.645\sqrt{\frac{1+2\sum_{i=1}^{s-1} {\text{ACF}_i}^2}{\hat{n}}},
\end{equation*}
where $\hat{n}$ is equal to $n$ and $n+h$ for the target and the reference series,
respectively. Non-seasonal series ($s=1$) and series where the number of observations is
fewer than three seasonal periods are not tested and not assumed as seasonal.

As some series may display multiplicative seasonality, the Box-Cox transformation
\citep{BC1964} is applied to each series before the STL \citep{Bergmeir2016-su}. The Box-Cox
transformation is defined as
\begin{equation*}
x_t'= \left\{
\begin{array}{ll}
      \log(x_t), & \lambda = 0, \\
      (x_t^\lambda -1)/\lambda,  & \lambda \neq 0,\\
\end{array}
\right.
\end{equation*}
where $x_t$ is a time series and $\lambda \in [0,1]$ is selected using the method of
\cite{Guerrero1993}, as implemented in the \textit{forecast} package for R
\citep{forecastR}. After Box-Cox transformation, $x_t'$ can be decomposed using STL method: $x_t' = T_t' + S_t' + R_t'$. To perform seasonal adjustment on the series $x_t$ with multiplicative seasonality, we first remove the seasonal component $S_t'$ from the Box-Cox transformed series $x_t'$, and denote the seasonal adjusted series as $x_{t, \mathrm{SA}}' = T_t' + R_t'$. Then the inverse Box-Cox transformation is applied to $x_{t, \mathrm{SA}}'$:
\begin{equation*}
x_{t, \mathrm{SA}}= \left\{
\begin{array}{ll}
      \exp(x_{t, \mathrm{SA}}'), & \lambda = 0, \\
      (\lambda x_{t, \mathrm{SA}}' + 1)^{(1/\lambda)},  & \lambda \neq 0,\\
\end{array}
\right.
\end{equation*}
where  $x_{t, \mathrm{SA}}$ is the final seasonal adjusted series of $x_t$.

As the forecasts produced by the seasonally adjusted series do not contain seasonal
information, we need to reseasonalise them with Step 7. Moreover, since the seasonal
component removed is Box-Cox transformed, the forecasts must also be transformed
using the same $\lambda$ calculated earlier. Having recovered the seasonality on the
transformed forecasts, a final inverse transformation is applied. As in STL decomposition the seasonal component changes over time, seasonality recovery is based on the latest available seasonal cycle. For instance, if the target series is of monthly frequency, then the last twelve estimated seasonal indices are used to reseasonalise the forecasts. If the forecast horizon is longer than the seasonal cycle, then these last estimated seasonal indices are re-used as many times needed.

\subsubsection{Smoothing}\label{sec:smoothing}

Smoothing is performed by utilizing the Loess method, as presented by \cite{Cleveland1992}
and implemented in the \textit{stats} package for R. In short, a local model is computed,
with the fit at point $t$ being the weighted average of the neighbourhood points and the
weights being proportional to the distances observed between the neighbours and point
$t$. Similarly to STL, Loess decomposes the series into the trend and remainder
components. Thus, by using the trend component, outliers and noise are effectively removed,
and it is easier to find similar series. Moreover, smoothing can help us obtain a more
representative forecast origin (last historical value of the series), potentially
improving forecasting accuracy \citep{SPILIOTIS201992}.

While we could directly use the smoothed trend component from STL in the previous step, we opt for a separate smoothing on the seasonally adjusted data (which consists of the trend and remainder components from STL). The reason for this is twofold. First, while a Box-Cox transformation is necessary before deseasonalising the data, as the seasonal pattern may be multiplicative and therefore impossible to be properly handled by STL, using the Box-Cox transformed smoothed trend component from STL would not allow us to correctly identify and match different trend patterns (such as additive versus multiplicative) between the target and the reference series. So, we separately smooth the seasonally-adjusted data, after an inverse Box-Cox transformation is applied to the sum of trend and remainder components from STL. Second, keeping the Loess smoothing separate to the deseasonalisation process allows for consistency across series that are identified as seasonal or not, as well as across frequencies of data.

\subsubsection{Scaling}\label{sec:scaling}

Scaling refers to translating the target and the reference series at the same magnitude so
that they are comparable to each other. This process can be done in various ways, such as
by dividing each value of a time series by a simple summary statistic (max, min, mean,
etc.), by restricting the values within a specific range (such as in $[0, 1]$), or by
applying a standard score. Since the forecast origin, the last historical value of the series, is the most crucial observation in
terms of forecasting, we divide each point by this specific value. A similar approach has
been successfully applied by \cite{Smyl2019-oy}. A different scaling needs to be considered to
avoid divisions by zero if either the target or the reference series contain zero
values. Finally, inverse scaling is applied to return to the target
series's original level with Step 6 once the forecasts have been produced. This is achieved via multiplying
each forecast by the origin.

\subsection{Similarity \& forecasting}\label{sec:similarity}

One disadvantage of forecasting using a statistical model is that a DGP is explicitly
assumed, although it might be difficult or even impossible to capture in
practice. Notwithstanding, our proposed methodology searches in a set of reference series
to identify similar patterns to those of the target series we need to forecast.

Given the preprocessed target series, $\tilde{y}$, and the $m$ preprocessed reference
series, $\tilde{Q}$, we search for similar series as follows: For each series, $i$, in the
reference set, $\tilde{Q}(i)$, we calculate the distance between its historical values,
$\tilde{Q}(i)_{1, \dots, n}$, and the ones of the target series using a distance
measure. The result of this process is a vector of length $m$ distances that correspond
to pairs of the target and the reference series available.

In terms of measuring distances, we consider three alternatives. The first one is the
$\mathcal{L}_1$ norm, which is equivalent to the sum of the absolute deviations between
$\tilde{y}$ and $\tilde{Q}(i)_{1, \dots, n}$. The second measure is the $\mathcal{L}_2$
norm (Euclidean distance), which is equivalent to the square root of the sum of the squared
deviations. The third alternative involves the utilization of the DTW. DTW can match sequences that are similar, but locally out of phase, by ``stretching'' and ``contracting'', and thus it allows non-linear mapping
between two time series. That is, $\tilde{y}_t$ can be matched either with $\tilde{Q}(i)_t$, as done with
$\mathcal{L}_1$ and $\mathcal{L}_2$, or with previous/following points of
$\tilde{Q}(i)_t$, even if these points have been already used in other matches. The three distance measures are formally expressed as
\begin{align*}
d_{\mathcal{L}_1}(\tilde{y}, \tilde{Q}(i)_{1, \dots, n})& = \norm{\tilde{y}_t-\tilde{Q}(i)_t}_1, \\
d_{\mathcal{L}_2}(\tilde{y}, \tilde{Q}(i)_{1, \dots, n}) &= \norm{\tilde{y}_t-\tilde{Q}(i)_t}_2, \\
d_\text{DTW}(\tilde{y}, \tilde{Q}(i)_{1, \dots, n}) &= D(n,n),
\end{align*}
\noindent where $D(n,n)$ is computed recursively as
\begin{equation}
    D(v,w) =  |\tilde{y}_v-\tilde{Q}(i)_w| + \min
\left\{
\begin{matrix}
D(\tilde{y}_v, \tilde{Q}(i)_{w-1})\\
D(\tilde{y}_{v-1}, \tilde{Q}(i)_{w-1})\\
D(\tilde{y}_{v-1}, \tilde{Q}(i)_w)\\
\end{matrix}
\right\}.
\label{eq:DTWrec}
\end{equation}
Equation (\ref{eq:DTWrec}) returns the total variation of two vectors,
$\tilde{y}_{1, \dots, v}$ and $\tilde{Q}(i)_{1, \dots, w}$. Note that DTW assumes a
mapping path from $(1,1)$ to $(n,n)$ with an initial condition of
$D(1,1) = |\tilde{y}_1-\tilde{Q}(i)_1|$.

The main differences among the three distance measures are: (1) DTW allows distortion in
the time axis, while $\mathcal L_1$ and $\mathcal L_2$ distances are more sensitive to
time distortion. Therefore, DTW introduces more flexibility to the process, allowing the
identification of similar series even when they display signal transformations such as
shifting and scaling, (2) allowing many-to-one point comparisons, DTW is more robust to
outliers or noise, (3) DTW can compare time series with different lengths, while the other
two measures are only applicable to time series with the same length, and (4) although DTW
is frequently chosen as the distance measure for time series related tasks such as
clustering and classification for its aforementioned merits, when dealing with large
datasets, DTW does not scale very well due to its quadratic time complexity. In contrast,
$\mathcal L_1$ and $\mathcal L_2$ distance measures are much easier to implement with
higher computational efficiency, making them also frequently used in a vast of time series
applications. We present the empirical differences among the three measures in the
proposed forecasting with cross-similarity approach in Section~\ref{sec:design}.

Having computed the distances between $\tilde{y}$ and $\tilde{Q}$, a subset of reference
series is chosen for aggregating their future paths and, therefore, forecasting the target
series. This is done by selecting the $k$ most similar series, i.e., the series that
display the smaller distances, as determined by the selected measure. In our experiment, we
consider different $k$ values to investigate the effect of pool size on
forecasting accuracy but demonstrate that any value higher than 100 is a suggested choice.

Essentially, we propose that the future paths from the most similar series can form the
basis for calculating the forecasts for the target series. Indeed, we do so by considering
the statistical aggregation of these future paths. The average is calculated for each
planning horizon.
This is an appealing approach in the sense that it does not involve
statistical forecasting in the traditional way: fitting statistical models and
extrapolating patterns. Instead, the real outcomes of a set of similar series are used to derive the forecasts.
We tested three averaging operators: the arithmetic mean, the median, and the weighted mean\footnote{The weighted mean is based on the degree of similarity: the values of the distances of the most similar series to the target series}. The median operator gave slightly better results than the two other operators, possibly due to its robustness and resistance to outliers. So, our empirical evaluation in section \ref{sec:evaluation} focuses on this operator.

\begin{figure}
    \centering
    \includegraphics[width=5in]{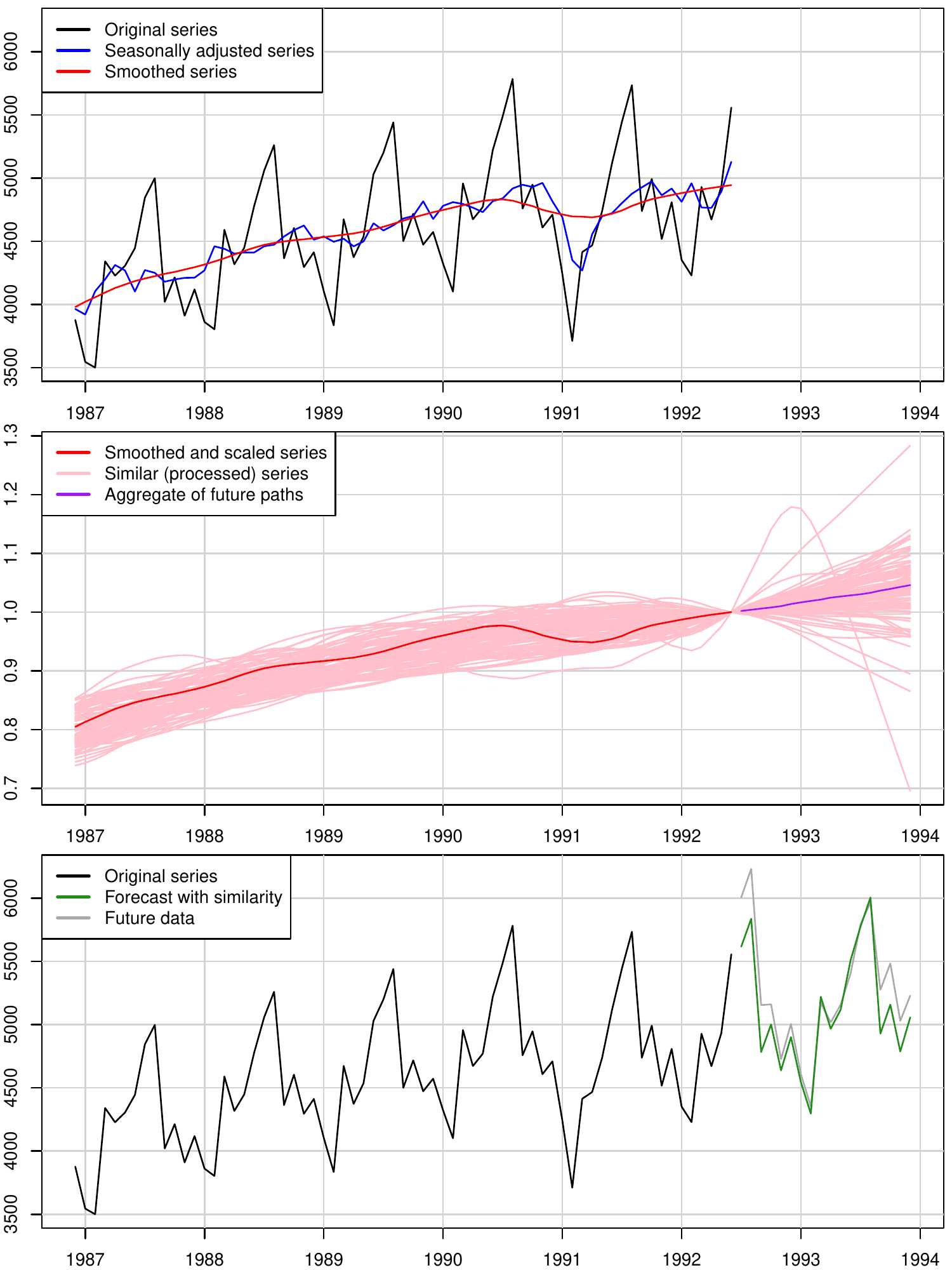}
    \caption{A toy example visualizing the methodology proposed for forecasting with
      cross-similarity. First, the target series is seasonally adjusted and smoothed (top
      panel). Then, the series is scaled, and similar reference series are used to
      determine its future path through aggregation (middle panel). Finally, the computed
      forecast is rescaled and reseasonalised to obtain the final forecast. The M495
      series of the M3 Competition data set is used as the target series. (For
      interpretation of the references to colour in this figure, the reader is referred to
      the web version of this article.)}
    \label{fig:toy}
\end{figure}

The proposed forecasting approach is demonstrated via a toy example, visualized in Figure
\ref{fig:toy}. The top panel presents the original target series, as well as the
seasonally adjusted and smoothed one. The middle panel shows the preprocessed
series (scaled values) together with the 100 most similar reference series used for
extrapolation. Finally, the bottom panel compares the rescaled and reseasonalised
forecasts to the actual future values of the target series.

Note that the above description assumes that Step 5 (forecasting and aggregation) is
completed before inverse scaling (Step 6) and recovering of seasonality (Step 7). Equally,
one could consider that Steps 6 and 7 are applied to each of the most similar reference
series, providing this way $k$ possible paths on the scale of the target series and
including the seasonal pattern identified in section \ref{sec:seasonality}. We denote
these rescaled and reseasonalised reference series as $\check{Q}_t$.  The aggregation of
these series would lead to the same point forecasts. Additionally, they can be used as the
basis for estimating the forecast uncertainty.

\subsection{Similarity \& prediction intervals}\label{sec:PI}

Time series forecasting uncertainty is usually quantified by prediction intervals, which
somehow depend on the forecastability of the target time series.  With a model-based
forecasting approach, although one could usually obtain a theoretical prediction interval,
the performance of such interval depends upon the length of the series, the accuracy of the model,
and the variability of model parameters. Alternatively, a straightforward attempt would be
bootstrapping the historical time series candidates and calculating the prediction
intervals based on their summary statistics \citep[e.g.,][]{thombs1990bootstrap,
  andre2002forecasting}. Such a procedure is model-dependent, which assumes that a known
model provides a promise fit to the data and requires specifying the distribution of the
error sequence associated with the model process.

Our interest is to find appropriate prediction intervals so that they could quantify the
uncertainty of the forecasts based on our similarity approach.  We use the variability
information from the rescaled and reseasonalised reference series, $\check{Q}_t$, as the
source of prediction interval bounds. However, we find that directly using the quantiles
or variance of the reference series may lead to lower-than-nominal coverage due to the
similarity (or low sampling variability) of reference series. To this end, we propose a
straightforward data-driven approach, in which the $(1-\alpha)100\%$ prediction interval
for a forecast $f_t$ is based on the a calibrated $\alpha/2$ and $1-\alpha/2$ quantiles of
the selected reference series $\check{Q}_t$ for the target $y_{t}$. The lower and upper
bounds for the prediction interval are defined as
\begin{equation}
  \label{eq:PI}
  L_t= (1-\delta)~F^{-1}_{\check{Q}_t}(\alpha/2) \mathrm{~and~} U_t= (1+\delta)~F^{-1}_{\check{Q}_t}(1-\alpha/2),
\end{equation}
respectively, where $F^{-1}_{\check{Q}_t}$ is the quantile based on the selected reference
series $\check{Q}_t$, and $\delta$ is a calibrating factor.

To evaluate the performance of the generated predictive
intervals,  we consider a scoring rule, the mean
scaled interval score (MSIS),  which is defined as
\begin{align}
\label{eq:msis}
\mathrm{MSIS} = \frac{1}{h}\frac{\sum_{t=n+1}^{n+h}(U_t-L_t)+\frac{2}{\alpha}(L_t-y_t)\bm{1}\left\{ y_t < L_t\right\} + \frac{2}{\alpha}(y_t - U_t)\bm{1}\left\{y_t>U_t\right\}}{\frac{1}{n-s}\sum_{t=s+1}^{n} \vert y_t-y_{t-s} \vert},
\end{align}
where $n$ is the historical length of the target time series, $s$ is the length of the seasonal period, and $h$ is the
forecasting horizon.  We aim to find an optimal calibrating factor $0 \leq \delta \leq 1$,
which minimizes the prediction uncertainty score (MSIS). To realize that, the target
series $y$ is first split into training and testing period, denoted as $y_{1, \dots, n-h}$
and $y_{n-h+1, \dots, n}$, respectively. We run the proposed forecasting approach to
$y_{1, \dots, n-h}$ and apply a grid search algorithm to search from a sequence of values
of $\delta \in \{0, 0.01, 0.02, \cdots, 1\}$ and find the optimal calibrating factor
$\delta^*$ that minimizes the MSIS values of the obtained prediction intervals of
$y_{1, \dots, n-h}$.  In the end, we get the prediction interval of $y$ by plugging the
optimal calibrating factor $\delta^*$ into Equation (\ref{eq:PI}).

\section{Evaluation} \label{sec:evaluation}

\subsection{Design}\label{sec:design}

In this paper, we aim to forecast the yearly, quarterly, and monthly series of the M1~\citep{Makridakis1982111} and M3
\citep{Makridakis2000a} forecasting competitions. These data sets have been widely used in the
forecasting literature with the corresponding research paper having been cited more than $1500$ and
$1600$ times, respectively, according to Google Scholar (as of 08/20/2020). The number of the yearly,
quarterly, and monthly series is presented in Table \ref{tab:data}, together with a
five-number summary of their lengths and the forecast horizon per frequency.

\begin{table}
\centering
\caption{The number of the target series, their lengths, and the forecasting horizon for each data frequency.}\vspace{0.25cm}
\begin{tabular}{ccccccccc}
\toprule
    \multirow{2}{2cm}{Frequency} &  \multirow{2}{1.5cm}{Number of series} & \multicolumn{5}{c}{Historical observations} & \multirow{2}{0.3cm}{$h$} \\
    & & Min & Q1 & Q2 & Q3 & Max & \\ \midrule
    Yearly & 826 & 9 & 14 & 17.5 & 26 & 52 & 6 \\
    Quarterly & 959 & 10 & 36 & 44 & 44 & 106 & 8 \\
    Monthly & 2045 & 30 & 54 & 108 & 116 & 132 & 18 \\ \midrule
    Total & 3830 \\
\bottomrule
\end{tabular}
\label{tab:data}
\end{table}

To assess the impact of the series length, we produce forecasts not only using
all the available history for each target series but also considering shorter historical
samples by truncating the long series and keeping the last few years of their
history. This is of particular interest in forecasting practice as in many enterprise
resource planning systems, such as SAP, only a limited number of years is usually
available. Table \ref{tab:data2} shows the cuts considered per frequency.

\begin{table}
\centering
\caption{The cuts of the target series considered.}\vspace{0.25cm}
\begin{tabular}{ccccccccc}
\toprule
  Frequency & \multicolumn{8}{c}{Up to (in years)} \\
  \midrule
  Yearly & 6 & 10 & 14 & 18 & 22 & 26 & 30 & 34  \\
  Quarterly & 3 & 4 & 5 & 6 & 7 & 8 & 9 & 10 \\
  Monthly & 3 & 4 & 5 & 6 & 7 & 8 & 9 & 10 \\
\bottomrule
\end{tabular}
\label{tab:data2}
\end{table}

For the purpose of forecasting based on similarity described in the previous section, we
need a rich and diverse enough set of reference series. For this purpose, we use the
yearly, quarterly, and monthly subsets of the M4 competition \citep{Makridakis2019-oy},
which consist of $23000$, $24000$, and $48000$ series, respectively. The lengths of these
series are, on average, higher than the lengths of the M1 and M3 competition data. The median
lengths are $29$, $88$, and $202$ for the yearly, quarterly, and monthly frequencies in M4,
respectively.

The point forecast accuracy is measured in terms of the Mean Absolute Scaled Error
\citep[MASE:][]{Hyndman06}. MASE is a scaled version of the mean absolute error, with the
scaling being the mean absolute error of the seasonal naive for the historical data. MASE
is widely accepted in the forecasting literature \citep[e.g.,][]{Franses2016-pj}. \cite{Makridakis2019-oy} also use this measure to evaluate
the point forecasts of the submitting entries for the M4 forecasting competition. Across
all horizons of a single series, the MASE value can be calculated as
\begin{equation*}
  \text{MASE} = \frac{1}{h} \frac{ \sum\nolimits_{t=n+1}^{n+h} {|y_{t}-f_{t}|} } {\frac{1}{n-s} \sum\nolimits_{t=s+1}^{n} |y_{t}-y_{t-s}|},
\end{equation*}
where $y_t$ and $f_t$ are the actual observation and the forecast for period $t$, $n$ is
the sample size, $s$ is the length of the seasonal period, and $h$ is the forecasting
horizon. Lower MASE values are better. Because MASE is scale-independent, averaging across
series is possible. We also have evaluated our approach using the mean absolute percentage error (MAPE). The results were consistent with the ones by MASE, and as such we do not provide the MAPE results in the manuscript for brevity.

To assess prediction intervals, we set $\alpha=0.05$ (corresponding to 95\% prediction intervals) and consider four measures --- MSIS, coverage, upper coverage and spread. MSIS is calculated as in Equation~\ref{eq:msis}. Coverage measures the percentage of times when the true values lie inside the prediction intervals. Upper coverage measures the percentage of times when the true values are not larger than the upper bounds of the prediction intervals: A proxy for achieved service levels. Spread refers to the mean difference of the upper and lower bounds scaled similarly to MSIS: A proxy for holding costs \citep{svetunkov2018old}. They are calculated as
\begin{align*}
\mathrm{Coverage}& = \frac{1}{h}\sum_{t=n+1}^{n+h}\bm{1}\left\{ y_t > L_t ~\&~ y_t < U_t\right\}, \\
\mathrm{Upper~coverage}& = \frac{1}{h}\sum_{t=n+1}^{n+h}\bm{1}\left\{ y_t < U_t\right\}, \\
\mathrm{Spread}& = \frac{\frac{1}{h}\sum_{t=n+1}^{n+h}(U_t-L_t)}{\frac{1}{n-s} \sum_{t=s+1}^{n} |y_{t}-y_{t-s}|},
\end{align*}
where $y_t$, $L_t$ and $U_t$ are the actual observation, the lower and upper bounds of the
corresponding prediction interval for period $t$, $n$ is the sample size, and $h$ is the
forecasting horizon. Note that the target values for the Coverage and Upper Coverage are
$95\%$ and $97.5\%$, respectively. Deviation from these values suggest under- or
over-coverage. Lower MSIS and Spread values are better.

\subsection{Investigating the performance of forecasting with cross-similarity}
\label{sec:design}

In this section, we focus on the performance of forecasting with cross-similarity and explore
the different settings, such as the choice of the distance measure, the pool size of
similar reference series (number of aggregates, $k$), as well as the effect of
preprocessing. Once the optimal settings are identified, in the next subsection, we
compare the performance of our proposition against that of four robust benchmarks for
different sizes of the historical sample.

Table \ref{tab:exploresimilarity} presents the MASE results of forecasting with
cross-similarity for each data frequency separately as well as across all frequencies
(Total). The summary across frequencies is a weighted average based on the series counts
for each frequency. Moreover, we present the results for each distance measure
($\mathcal{L}_1$, $\mathcal{L}_2$, and DTW) in rows and various values of $k$ in columns.

\begin{table}
\centering
\caption{The MASE performance of the forecasting with cross-similarity approach for different distance measures and pool sizes of similar reference series ($k$).}
\begin{tabular}{ccccccccc}
\toprule
\multirow{2}{*}{Frequency} & \multirow{2}{*}{Distance Measure} & \multicolumn{7}{c}{Number of aggregated reference series ($k$)} \\
&&1&5&10&50&100&500&1000 \\
\midrule
\multirow{3}{*}{Yearly}&$\mathcal{L}_1$  & 3.375 & 2.936 & 2.884 & 2.801 & 2.784 & 2.785 & 2.798 \\
&$\mathcal{L}_2$ & 3.378 & 2.960 & 2.876 & 2.813 & 2.800 & 2.794 & 2.805 \\
&$\text{DTW}$ & 3.345 & 2.948 & 2.846 & 2.781 & 2.777 & 2.783 & 2.805 \\
\midrule
\multirow{3}{*}{Quarterly}&$\mathcal{L}_1$ & 1.468 & 1.345 & 1.316 & 1.279 & 1.273 & 1.262 & 1.260 \\
&$\mathcal{L}_2$ & 1.488 & 1.335 & 1.305 & 1.278 & 1.273 & 1.261 & 1.261 \\
&$\text{DTW}$ & 1.440 & 1.316 & 1.297 & 1.257 & 1.254 & 1.250 & 1.250 \\
\midrule
\multirow{3}{*}{Monthly}&$\mathcal{L}_1$& 1.082 & 0.992 & 0.964 & 0.948 & 0.946 & 0.943 & 0.943 \\
&$\mathcal{L}_2$& 1.088 & 0.993 & 0.970 & 0.948 & 0.945 & 0.942 & 0.943 \\
&$\text{DTW}$ & 1.080 & 0.971 & 0.950 & 0.935 & 0.936 & 0.932 & 0.932 \\
\midrule
\multirow{3}{*}{Total}&$\mathcal{L}_1$& 1.673 & 1.500 & 1.466 & 1.431 & 1.424 & 1.420 & 1.422 \\
&$\mathcal{L}_2$& 1.682 & 1.503 & 1.465 & 1.433 & 1.427 & 1.421 & 1.424 \\
&$\text{DTW}$ & 1.659 & 1.484 & 1.446 & 1.414 & 1.413 & 1.411 & 1.416 \\
\bottomrule
\end{tabular}
\label{tab:exploresimilarity}
\end{table}

A comparison across the different values for the number of reference series, $k$, suggests
that large pools of representative series provide better performance. At the same time,
the improvements seem to tapper off when $k>100$. Based on the reference set we use in
this study, we identify a sweet point at $k=500$. The analysis presented in
section~\ref{sec:similarity-versus-model-based} focuses on this aggregate size. In any
case, we find that both the reference series's size and its similarity with the
target series affect the selection of the value of $k$.

\begin{figure}
    \centering
    \includegraphics[width=\textwidth]{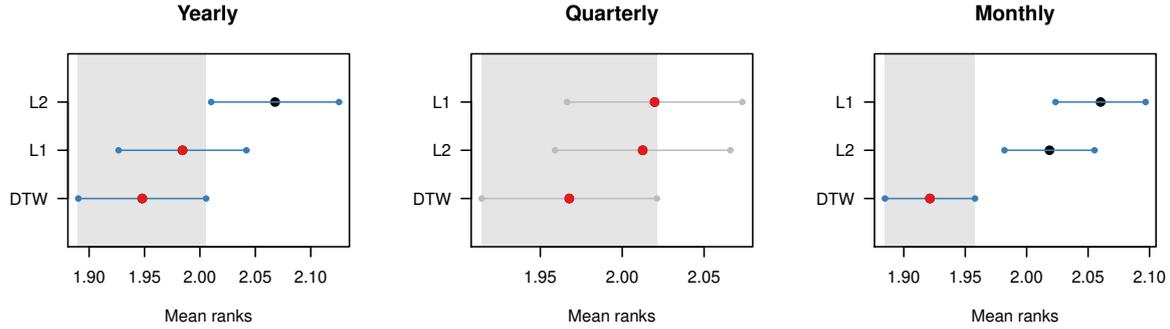}
    \caption{MCB significance tests for the three distance measures for each data frequency.}
    \label{fig:MCB1}
\end{figure}

Table \ref{tab:exploresimilarity} also shows that $\mathcal{L}_1$ and $\mathcal{L}_2$
perform almost indistinguishable across all frequencies. DTW almost always outperforms the
other two distance measures. However, the differences are small, to the degree of
$10^{-2}$ in our study. Given that the DTW is more computationally intensive than
$\mathcal{L}_1$ and $\mathcal{L}_2$ (approximately $\times6$, $\times10$, and {$\times27$}
for yearly, quarterly, and monthly frequencies, respectively), we further investigate the
statistical significance of the achieved performance improvements.  To this end, we apply
the Multiple Comparisons with the Best (MCB) test that compares whether the average
(across series) ranking of each distance measure is significantly different than the others (for
more details on the MCB, please see \cite{Koning2005}). With MCB, when the confidence
intervals of two methods overlap, their ranked performances are not statistically
different. The analysis is done for $k=500$. The results are presented in Figure
\ref{fig:MCB1}. We observe that DTW results in the best-ranked performance, which is statistically different from that of the other two distance measures only for the monthly frequency. We
argue that if the computational cost is a concern, one may choose between $\mathcal{L}_1$
and $\mathcal{L}_2$. Otherwise, DTW is preferable, both in terms of average forecast
accuracy and mean ranks. In the analysis below, we focus on the DTW distance measure.

The aforementioned results are based on the application of preprocessing (as described in section
\ref{sec:preprocessing}), including seasonal adjustment and smoothing, before searching
for similar series. Now we investigate the improvements in seasonal adjustment and
smoothing. In the Loess method used for smoothing, the parameter  ``span'' controls the degree of smoothing, which is set to $h$ in Table~\ref{tab:exploresimilarity}. To investigate how the degree of smoothing in the Loess method influences the accuracies of forecasting with cross-similarity, we consider 30\% less and 30\% more smoothing.  Note that the scaling process (as described
in section \ref{sec:scaling}) is always applied to make the target and reference series
comparable. Table \ref{tab:explorepreprocessing} presents the MASE results for DTW across
different $k$ values with and without the preprocessing described in sections
\ref{sec:seasonality} and \ref{sec:smoothing}, and with different amounts of smoothing.  The main findings are: (1) preprocessing
always provides better accuracy, so it is recommended with the forecasting
with cross-similarity approach, and (2) for yearly and quarterly data, which are usually smooth and relatively short, less smoothing is preferred, while monthly data prefer more smoothing. Therefore, we use 30\% less smoothing for yearly and quarterly data, and 30\% more smoothing for monthly data in the following sections of the manuscript. Note that the percentage ``30\%'' are arbitrarily selected here to demonstrate that smoothing improves forecasting if properly applied, and that other parameters could be used instead, possibly leading to even better results.

\begin{table}
\centering
\caption{The MASE performance of forecasting with cross-similarity, with and without seasonal
  adjustment and smoothing. The DTW distance measure is considered. ``span'' controls the degree of smoothing in the Loess method. A larger value of ``span'' means more smoothing. $h$ is the forecasting horizon.}
\begin{tabular}{clccccccc}
\toprule
\multirow{2}{*}{Frequency} & Seasonal adjustment & \multicolumn{7}{c}{Number of aggregated reference series ($k$)} \\
& and smoothing &1&5&10&50&100&500&1000 \\
\midrule
\multirow{4}{*}{Yearly}&NO & 3.649 & 2.967 & 2.898 & 2.823 & 2.819 & 2.828 & 2.845 \\
&YES, $\mathrm{span} = h \times 0.7$& 3.474 & 2.915 & 2.859 & 2.778 & 2.770 & 2.774 & 2.796 \\
&YES, $\mathrm{span} = h$& 3.345 & 2.948 & 2.846 & 2.781 & 2.777 & 2.783 & 2.805 \\
&YES, $\mathrm{span} = h \times 1.3$ & 3.252 & 2.901 & 2.849 & 2.824 & 2.808 & 2.820 & 2.842 \\
  \midrule
\multirow{4}{*}{Quarterly}&NO  & 1.734 & 1.508 & 1.457 & 1.471 & 1.484 & 1.504 & 1.507 \\
&YES, $\mathrm{span} = h \times 0.7$  & 1.481 & 1.319 & 1.274 & 1.246 & 1.247 & 1.246 & 1.246 \\
&YES, $\mathrm{span} = h$ & 1.440 & 1.316 & 1.297 & 1.257 & 1.254 & 1.250 & 1.250 \\
&YES, $\mathrm{span} = h \times 1.3$ & 1.446 & 1.341 & 1.313 & 1.276 & 1.279 & 1.277 & 1.274 \\
  \midrule
\multirow{4}{*}{Monthly}&NO & 1.381 & 1.190 & 1.130 & 1.122 & 1.126 & 1.153 & 1.171 \\
&YES, $\mathrm{span} = h \times 0.7$ & 1.159 & 1.002 & 0.977 & 0.959 & 0.958 & 0.958 & 0.959 \\
&YES, $\mathrm{span} = h$ & 1.080 & 0.971 & 0.950 & 0.935 & 0.936 & 0.932 & 0.932 \\
&YES, $\mathrm{span} = h \times 1.3$ & 1.028 & 0.955 & 0.933 & 0.926 & 0.926 & 0.922 & 0.923 \\
 \bottomrule
\end{tabular}
\label{tab:explorepreprocessing}
\end{table}

\subsection{Similarity versus model-based forecasts}\label{sec:similarity-versus-model-based}

Having identified the optimal settings (DTW, $k=500$, and preprocessing) for forecasting
with cross-similarity, abbreviated from now on simply as `Similarity', in this subsection we
turn our attention to comparing the accuracy of our approach against well-known
forecasting benchmarks. We use four benchmark methods. The forecasts with the first method
derive from the optimally selected exponential smoothing model when applying selection
with the corrected (for small sample sizes) Akakie's Information Criterion
($\text{AIC}_c$). This optimal selection occurs per series individually so that a different
optimal model may be selected for different series. We use the implementation available in
the \textit{forecast} package for the R statistical software, and in particular the
\texttt{ets()} function \citep{Hyndman2008b}. The second benchmark is the automatically selected most appropriate autoregressive integrated moving average (ARIMA) model, using the implementation of the \texttt{auto.arima()} function \citep{Hyndman2008b}. The third benchmark is the Theta method \citep{Assimakopoulos2000521}, which was the top performing method in the M3 forecasting competition \citep{Makridakis2000a}. Finally, the last benchmark is the simple
(equally-weighted) combination of three exponential smoothing models: Simple Exponential
Smoothing, Holt's linear trend Exponential Smoothing, and Damped trend Exponential
Smoothing. This combination is applied to the seasonally adjusted data (multiplicative
classical decomposition) if the data have seasonal patterns with the seasonality test
described in section \ref{sec:seasonality}. This combination approach has been used as a
benchmark in international forecasting competitions
\citep{Makridakis2000a,Makridakis2019-oy} and it is usually abbreviated as SHD.

We have also tested the performance of a self-similarity approach through $k$NN ($k$-Nearest-Neighbour) for time series, implemented in the \textit{tsfknn} R package \citep{Martinez2019-ve}, and found that focusing merely on similar patterns within a series results in very poor forecasting performance. More importantly, \textit{tsfknn} is not applicable when historical information is limited. As such, we decide not to include this approach as a benchmark in our study.

\begin{figure}
    \centering
    \includegraphics[width=4.2in]{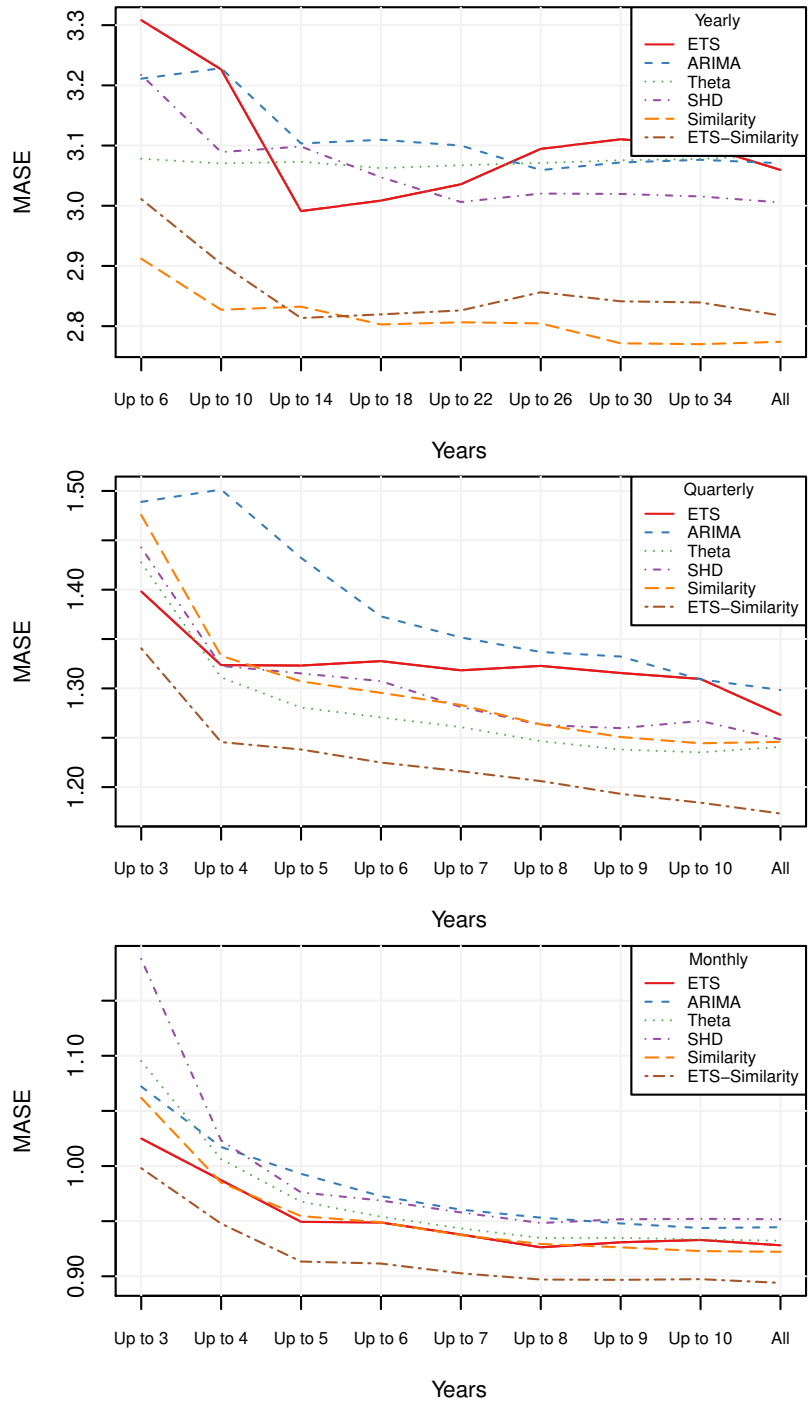}
    \caption{Benchmarking the performance of Similarity against ETS, ARIMA, Theta, and SHD for various
      historical sample sizes. (For interpretation of the references to colour in this
      figure, the reader is referred to the web version of this article.)}
    \label{fig:benchmarks}
\end{figure}

Figure \ref{fig:benchmarks} shows the accuracy of our Similarity approach against the four benchmarks,
ETS, ARIMA, Theta and SHD. The comparison is made for various historical sample sizes to examine the
effect of data availability. We observe:
\begin{itemize}[noitemsep,nolistsep]
\item In the yearly frequency, Similarity always outperforms the four benchmarks regardless
  of the length of the available history. It is worth mentioning that ETS improves when
  not all available observations are used for model fitting (truncated target
  series). Using just the last $14$ years of the historical samples gives the best
  accuracy in the yearly frequency for ETS. ARIMA, SHD, and Similarity perform better when more
  data are available. Theta is not affected by the length of the series for the yearly frequency.
\item In the quarterly frequency, similarity performs very competitively against the statistical benchmarks when the length of the series is longer than four years. Only Theta achieves, on average, better performance than similarity. The performance of all methods is improved as the number of observations increases.
\item In the monthly frequency, the performance of ETS and Similarity is indistinguishable, outperforming all other statistical benchmarks. Lengthier monthly series generally result in improved performance up to a point: if more than $7$ or $8$ years of data are available, then the changes in forecasting accuracy are small.
\end{itemize}

Figure \ref{fig:benchmarks} also shows the performance of the simple forecast combination
of ETS and Similarity (``ETS-Similarity'')\footnote{Other simple combinations of ARIMA, Theta, and SHD
  with Similarity were also tested, having on average same or worse performance to the
  ETS-Similarity simple combination.}, which takes their arithmetic mean as the final forecasts. The argument is that these two forecasting
approaches are diverse in nature (model-based versus data-centric) but also robust when
applied separately. So we expect that their combination will also perform well
\citep{Lichtendahl2019}. We observe that this simple combination performs on par to
Similarity for the yearly frequency, being much better than any other approach at the
seasonal frequencies. Overall, the simple combination of ETS-Similarity is the best
approach. This suggests that there are different benefits in terms of forecasting
performance improvements with both model-based and data-centric approaches. Solely
focusing on one or the other might not be ideal.

\begin{figure}
    \centering
    \includegraphics[width=\textwidth]{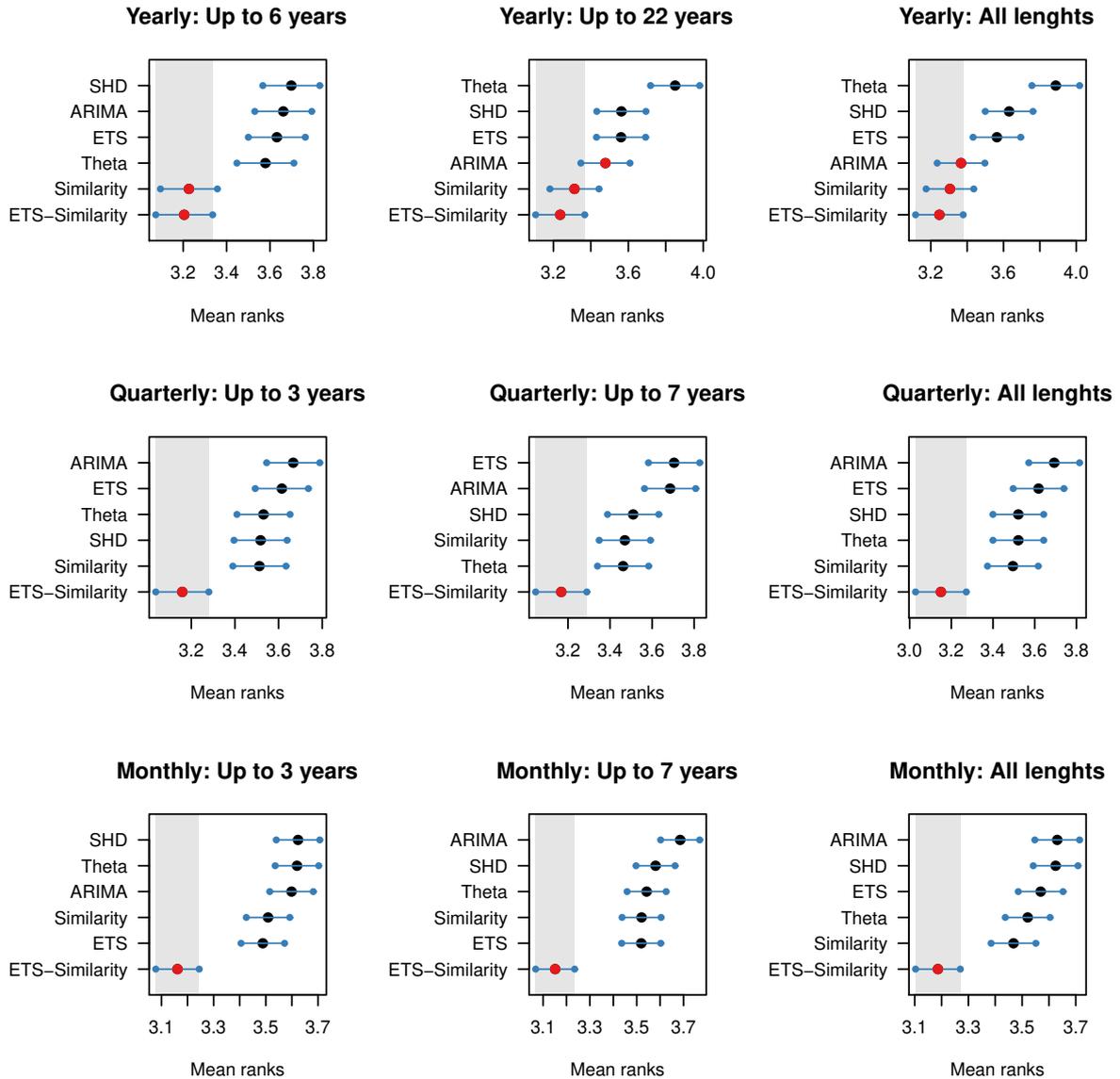}
    \caption{MCB significance tests for ETS, ARIMA, Theta, SHD, Similarity, and ETS-Similarity for each
      data frequency and various sample sizes.}
    \label{fig:MCB2}
\end{figure}

Finally, we compare the differences in the ranked performance of the five approaches
(ETS, ARIMA, Theta, SHD, and Similarity) and the one combination (ETS-Similarity) in terms of their
statistical significance (MCB). The results are presented in the nine panels of Figure
\ref{fig:MCB2} for each frequency (in rows) and short, medium, and long historical samples
(in columns). We observe:
\begin{itemize}[noitemsep,nolistsep]
\item Similarity is significantly better than the statistical benchmarks for the short yearly
  series. At the same time, similarity performs statistically similar to the best of the statistical benchmarks for other lengths and frequencies.
\item A simple combination of ETS and Similarity is always ranked
  1\textsuperscript{st}. Moreover, its performance is significantly better
  compared to ETS, Theta, and SHD for all frequencies and historical sample sizes (their intervals
  do not overlap). ARIMA, Similarity, and ETS-Similarity are not statistically different at the
  yearly frequency, but the combination approach is better at the seasonal data.
\end{itemize}

\subsection{Evaluating uncertainty estimation}\label{sec:evalPI}

We firstly investigate the importance of the calibrating procedure of prediction intervals
by exploring the relationship between the forecastability of the target series and the selected calibrating factor
$\delta^*$. We follow \citet{Kang2017345} and use the spectral entropy to measure the
``forecastability'' of a time series as
\begin{align*}
  \mathrm{Forecastability} = 1 + \int_{-\pi}^{\pi} \hat f_y(\gamma) \log \hat f_y(\gamma) \mathrm{d} \gamma,
\end{align*}
where $\hat f_x(\gamma)$ is an estimate of the spectrum of the time series that describes
the importance of frequency $\gamma$ within the period domain of  a given time series $y$. A larger value of
Forecastability suggests that the time series contains more signal and is easier to forecast.  On the other hand, a smaller value of forecastability indicates more
uncertainty about the future, which suggests that the time series is harder to forecast.

\begin{figure}
    \centering
    \includegraphics[width=0.8\textwidth]{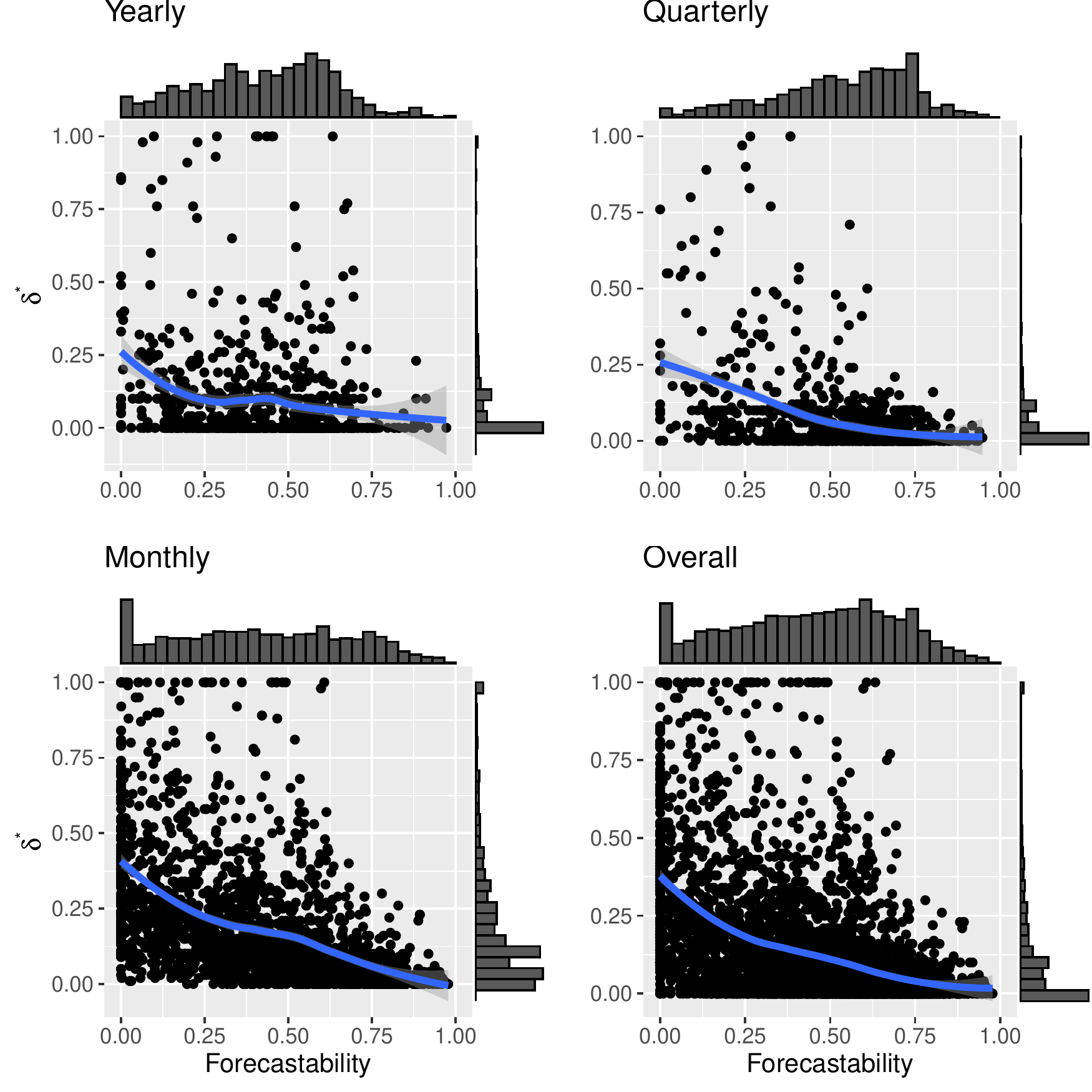}
    \caption{Relationship between forecastability and the optimal calibrating factor
      ($\delta$) using a nonparametric Loess regression curve (blue line) for yearly (top
      left), quarterly (top right), monthly (bottom left) and overall (bottom right) data. The top and right margins of each subplot are the histograms of
      forecastability and the optimal calibrating factor $\delta^*$, respectively. }
    \label{fig:deltaVSforecastability}
\end{figure}

Figure \ref{fig:deltaVSforecastability} depicts the relationship between forecastability
and $\delta^*$ for the studied time series by showing the scatter plots of the aforementioned variables for yearly, quarterly, and monthly data, as well as the complete dataset. The corresponding nonparametric loess regression curves are also shown. Along the top and right margins of each scatter plot, we show the histograms of forecastability and $\delta^*$ to present their distributions.
From Figure \ref{fig:deltaVSforecastability}, we find that time series with lower forecastability values yield higher calibrating
factors $\delta^*$. That is, to obtain a more appropriate prediction interval, we need to calibrate more for time series that are harder to forecast. The forecastability of a large proportion of the monthly data is weak when compared to that of the yearly and quarterly data, which
makes the overall dataset hard to forecast. The nonparametric loess regression curves
indicate that there is a strong dependence between forecastability and the calibration factor,
which is strong evidence of elaborating a calibrating factor in the prediction intervals
for hard-to-forecast time series.

We proceed by comparing the forecasting performances based on the calibrated prediction intervals of
Similarity and other benchmarks. Table \ref{table:PIevaluation} shows the performance of
Similarity against the four benchmarks, ETS, ARIMA, Theta, and SHD, regarding prediction intervals. The
performance of the forecast combination of ETS and Similarity (ETS-Similarity) is also
shown. Our findings are as follows:

\begin{itemize}
\item For yearly data, similarity significantly outperforms the four benchmarks according to MSIS, while also providing higher
  coverage and upper coverage. The simple combination of ETS and Similarity achieves similar performance with similarity, with higher coverage and tighter prediction intervals. Overall, we
  conclude that similarity significantly outperforms ETS, ARIMA, Theta, and SHD for yearly data.
\item For quarterly and monthly data, similarity displays similar performance to that of ETS. However, it yields significantly higher upper coverage and at the same time loses some spread. The simple combination of ETS and Similarity achieves
  the best performances regarding MSIS and (upper) coverage levels compared with the four benchmarks.
\end{itemize}

\begin{table}
\caption{Benchmarking the performance of Similarity against ETS, ARIMA, Theta, SHD, and ETS-Similarity with regard to MSIS, coverage, upper coverage and spread of prediction intervals.}
\label{table:PIevaluation}
\centering
\begin{tabular}{lcccc}
  \toprule
  & MSIS & Coverage (\%) & Upper coverage (\%) & Spread \\
  &  & Target: 95\% & Target: 97.5\% &  \\
  \midrule
  & \multicolumn{4}{c}{Yearly}\\
  ETS           & 37.008 & 81.578 & 86.844 & 11.967 \\
  ARIMA         & 45.590 & 77.260 & 86.077 & \textbf{8.364} \\
  Theta         & 39.568 & 80.851 & 84.705 & 8.871 \\
  SHD           & 42.424 & 77.220 & 83.051 & 8.506 \\
  Similarity    & \textbf{26.432} & 88.680 & \textbf{94.592} & 13.567 \\
  ETS-Similarity& 26.809 & \textbf{89.588} & 93.119 & 12.767 \\

  \midrule
   & \multicolumn{4}{c}{Quarterly}\\
  ETS           & 12.961 & 85.076 & 91.489 & 4.805 \\
  ARIMA         & 14.982 & 80.214 & 91.919 & \textbf{4.173} \\
  Theta         & 13.785 & 84.541 & 90.667 & 4.309 \\
  SHD           & 13.409 & 84.333 & 90.302 & 4.391 \\
  Similarity    & 12.823 & 86.861 & 94.121 & 5.778 \\
  ETS-Similarity& \textbf{11.245} & \textbf{89.937} & \textbf{94.799} & 5.292 \\

  \midrule
   & \multicolumn{4}{c}{Monthly}\\
  ETS           & 7.333 & 90.685 & 94.224 & 4.300 \\
  ARIMA         & 8.348 & 89.343 & 94.659 & 4.087 \\
  Theta         & 7.984 & 88.840 & 93.371 & \textbf{4.072} \\
  SHD           & 7.895 & 89.343 & 93.461 & 4.276 \\
  Similarity    & 7.643 & 90.497 & 95.873 & 4.853 \\
  ETS-Similarity& \textbf{6.591} & \textbf{93.146} & \textbf{96.438} & 4.576 \\
  \bottomrule
\end{tabular}
\end{table}

\section{Discussions} \label{sec:discussions}

Statistical time series forecasting typically involves selecting or combining the most
accurate forecasting model(s) per series, which is a complicated task significantly
affected by data, model, and parameter uncertainties. On the other hand, nowadays, big data
allows forecasters to improve forecasting accuracy through cross-learning, i.e., by
extracting information from multiple series of similar characteristics. This practice has
been proved highly promising, primarily through the exploitation of advanced machine
learning algorithms and fast computers \citep{Makridakis2019-oy}. Our results confirm that
data-centric solutions offer a handful of advantages over traditional model-based ones,
relaxing the assumptions made by the models, while also allowing for more
flexibility. Thus, we believe that extending forecasting from within series to across
series, is a promising direction to forecasting.

An important advancement of our forecasting approach over other cross-learning ones, is
that similarity derives directly from the data, not depending on the extraction of a
feature vector that indirectly summarizes the characteristics of the series
\citep{Petropoulos2014152, Kang2017345, Kang2019-ar}. To this end, the uncertainty related
to the choice and definition of the features used for matching the target to the reference
series is effectively mitigated. Moreover, no explicit rules are required for determining
what kind of statistical forecasting model(s) should be used per case
\citep{Monteros2019-oy}. Instead of specifying a pool of forecasting models and an
algorithm for assigning these models to the series, a distance measure is defined and
exploited for evaluating similarity. Finally, forecasting models are replaced by the actual
future paths of the similar reference series.

Our results are significant for the practice of business research with more accurate forecasts translating into better business decisions. Forecasting
is an important driver for reducing inventory associated costs and waste in supply chains
\citep[for a comprehensive review on supply chain forecasting,
see][]{Syntetos2016-ew}. Small improvements in forecast accuracy are usually amplified in
terms of the inventory utility, namely inventory holding and achieved target service
levels \citep{Syntetos2010,syntetos2015forecasting}. At the same time, forecast accuracy is also essential to
other areas of business research, such as humanitarian operations and logistics
\citep{Rodriguez-Espindola2018-mr}, marketing \citep{qian2014using}, and finance
\citep{YU2019}.

While the point forecasts are oftentimes directly used in inventory settings, we show that
forecasting with cross-similarity allows for better estimation of the forecast uncertainty compared to
statistical benchmarks. The upper coverage rates of our approach are superior to that of statistical approaches,
directly pointing to higher achieved customer service levels. This is achieved by a minimal increase
of the average spread of the prediction intervals,
suggesting a small difference in the corresponding holding cost.

Our study also has implications for software providers of forecasting support systems. We
offer our code as an open-source solution together with a web interface\footnote{Available
  here: \url{https://fotpetr.shinyapps.io/similarity/}} (developed in R and Shiny) where a
target series can be forecasted through similarity, as described in section
\ref{sec:methodology}, using the large M4 competition data set as the reference set. We
argue that our approach is straightforward to implement based on existing solutions,
offering a competitive alternative to traditional statistical modelling. Forecasting with
cross-similarity can expand the existing toolboxes of forecasting software. Given that none
approach is the best for all cases, a selection framework (such as time series
cross-validation) can optimally pick between statistical models or forecasting with
cross-similarity based on past forecasting performance.

However, the computational time is a critical factor that should be carefully taken into
consideration, especially when forecasting massive data collections. This is particularly
true in supply chain management, where millions of item-level forecasts must be produced on
a daily basis \citep{SEAMAN2018822}. An advantage of our approach is that the
computational tasks in forecasting with cross-similarity can be easily programmed in parallel
compared to multivariate models. Moreover, since the DTW distance measure is more
computationally intensive than the two other measures presented in this study, an option
would be to select between them based on the results of an ABC-XYZ analysis
\citep{RAMANATHAN2006695}. This analysis is based on the Pareto principle (the 80/20
rule), i.e., the expectation that the minority of cases has a disproportional impact on the
whole. In this respect, the target series could be first classified as A, B, or C,
according to their importance/cost, and as X, Y, or Z, based on how difficult it is to be
accurately forecasted. Then, series in the AZ class (important but difficult to forecast)
could be predicted using DTW, while the rest using another, less computationally intensive
distance measure.

Forecasting with cross-similarity is based on the availability of a rich collection of reference
series. In order to have appealing forecasting performance, such a reference dataset
should be as representative (see \citet{Kang2019-ar} for a more rigorous definition) as
possible to the target series, which is easy to achieve in business cycles because of data
accumulation. To illustrate and empirically demonstrate the effectiveness of the approach,
we used the M4 competition data set as a reference. This data set is considered to
represent the reality appropriately \citep{Spiliotis2019M3}. However, if our approach is
to be applied to the data of a specific company or sector, then it would make sense that
the reference set is derived from data of that company/sector so as to be as
representative as possible. In the case that it is challenging to identify appropriate
reference series for the target series, then generating series with the desirable
characteristics \citep{Kang2019-ar} is an option.

We have empirically tested our approach on three representative data frequencies: yearly,
quarterly, and monthly. We have no reason to believe that our approach would not perform
well for higher frequency data, such as weekly, daily, or hourly. If multiple seasonal
patterns appear, as it could be the case for the hourly frequency with periodicity within
a day (every 24 hours) and within a week (every 168 hours), then a multiple seasonal
decomposition needs to be applied instead of the standard STL (the \textit{forecast}
package for R offers the \texttt{mstl()} function for this purpose). On the other hand,
our approach is not suitable as-is for intermittent demand data, where the demand values
for several periods are equal to zero. In this case, one could try forecasting with
cross-similarity without applying data preprocessing. A similar approach was proposed by
\cite{Nikolopoulos2016-gr} who focused on identifying patterns within intermittent demand
series rather than across series.

\section{Concluding remarks} \label{sec:conclusions}

In this paper, we introduce a new forecasting approach that uses the future paths of
similar reference series to forecast a target series. The advantages of our proposition are
that it is model-free, in the sense that it does not rely on statistical forecasting
models, and, as a result, it does not assume an explicit DGP. Instead, we argue that
history repeats itself (d\'ej\`a vu) and that the current data patterns will resemble the
patterns of other already observed series. The proposed approach is data-centric and
relies on the availability of a rich, representative reference set of series -- a not so
unreasonable requirement in the era of big data.

We examined the performance of the new approach on a widely-used data set and benchmarked
it against four robust forecasting methods, namely the automatic selection of the best
model from the Exponential Smoothing family (ETS), as well as the ARIMA family, the Theta method, and the equal-weighted combination of
Simple, Holt, and Damped exponential smoothing (SHD). We find that in most frequencies,
the new approach is more accurate than the benchmarks. Moreover, forecasting with cross-similarity
is able to better estimate the uncertainty of the forecasts, resulting in better upper coverage
levels, which are crucial for fulfilling customer demand. Finally, we propose a simple combination
of model-based and model-free forecasts, which results in an accuracy that is always significantly
better than the one or the other separately.

The innovative proposition of forecasting with cross-similarity and without models points
towards several future research paths. For example, in this study we do not
differentiate the reference series to match the industry/field of the target series. It
would be interesting to explore if such matching would further improve the accuracy of
forecasting with cross-similarity.

\section*{Acknowledgements}


Yanfei Kang is supported by the National Natural Science Foundation of China
(No. 11701022) and the National Key Research and Development Program
(No. 2019YFB1404600). Feng Li is supported by the National Natural Science Foundation of
China (No. 11501587) and the Beijing Universities Advanced Disciplines Initiative
(No. GJJ2019163).


\bibliographystyle{model5-names}
\bibliography{My_Collection}

\end{document}